\documentclass[fleqn,10pt]{wlscirep}
\title{Grating chips for quantum technologies}

\author{James P.\ McGilligan}
\author{Paul F.\ Griffin}
\author{Rachel Elvin}
\author{Stuart J.\ Ingleby}
\author{Erling Riis}
\author[1*]{Aidan S.\ Arnold}
\affil{Department of Physics, SUPA, University of Strathclyde, Glasgow, G4 0NG, United Kingdom}
\usepackage{hyperref}
\newcommand{\doi}[2]{\href{http://dx.doi.org/#1}{#2}}

\affil[*]{aidan.arnold@strath.ac.uk}


\begin{abstract}
We have laser cooled $3\times10^6$ $^{87}$Rb atoms to 3~$\mu$K in a micro-fabricated grating magneto-optical trap (GMOT), enabling future mass-deployment in highly accurate compact quantum sensors. We magnetically trap the atoms, and use Larmor spin precession for magnetic sensing in the vicinity of the atomic sample. Finally, we demonstrate an array of magneto-optical traps  with a single laser beam, which will be utilised for future cold atom gradiometry. 
\end{abstract}

\begin{document}

\flushbottom
\maketitle
%
%

\section*{Introduction}

Laser cooling has led to profound advances in the field of metrology due to the increased interrogation times gained from the low velocity ensembles \cite{katori, ludlow}. The longer interrogation times correspond to narrower transition resonances that have proven critical to the field of quantum sensors. The issue hindering the use of cold atoms as the basis of a portable quantum device is the large spatial scale of the apparatus \cite{kitching1, syrte1, rolsten1}. Significant efforts have been made in recent years to overcome this constraint by developing a compact technology that facilitates ultra-cold atomic ensembles as a portable sensing device \cite{zhang}.

Attempts to achieve a scalable cooling apparatus have been demonstrated in the past in the form of the pyramid magneto-optical trap, PMOT \cite{pollock1,pyramid1,pyramid2}. This consists of a structure with four highly reflective walls, angled to overlap the reflected beams with the incident light. The device needs to be maintained in the vacuum, due to the MOT forming within the pyramid volume. Further difficulties arise when imaging these atoms due to optical restrictions caused by the pyramid geometry. An alternative tetrahedral mirror pyramid geometry \cite{vangeleyn1} ameliorates all negative features of the standard PMOT.

In this paper we build upon our work with the grating magneto-optical trap, GMOT \cite{mcgilligan1,mcgilligan2,mcgilligan3, nshii}, and show it is a viable platform for compact quantum sensing by generating a truly ultra-cold atomic sample with a micro-fabricated chip. This study also focusses on magnetically trapping a cold sample for measurement of the background Rb pressure present in the vacuum chamber. Furthermore, we utilise Larmor spin precession for magnetometry in order to null stray static and gradient fields in the lab environment. Finally, we demonstrate an array of cold atoms generated above a diffractive optic with a single laser, with an outlook to demonstrating a compact, cold atom gradiometer.

\section*{The grating magneto-optical trap}

The GMOT cools atoms by a balanced radiation pressure between a single incident laser beam and the diffracted orders generated from the grating surface. Recent investigations of physical geometry have aimed at the optimisation of the diffractive optics for generation of a balanced radiation pressure, with the current conclusion that a 1200~nm period of 60:40 duty cycle in a triangular geometry performs best when used outside the vacuum \cite{mcgilligan1}. In addition to the laser alignment and intensity balance controlled by the diffractive optic \cite{mcgilligan1}, the critical parameter to achieve sub-Doppler temperature is the magnetic field.

\begin{figure}[t]
\centering
\centering
\includegraphics[width=14cm]{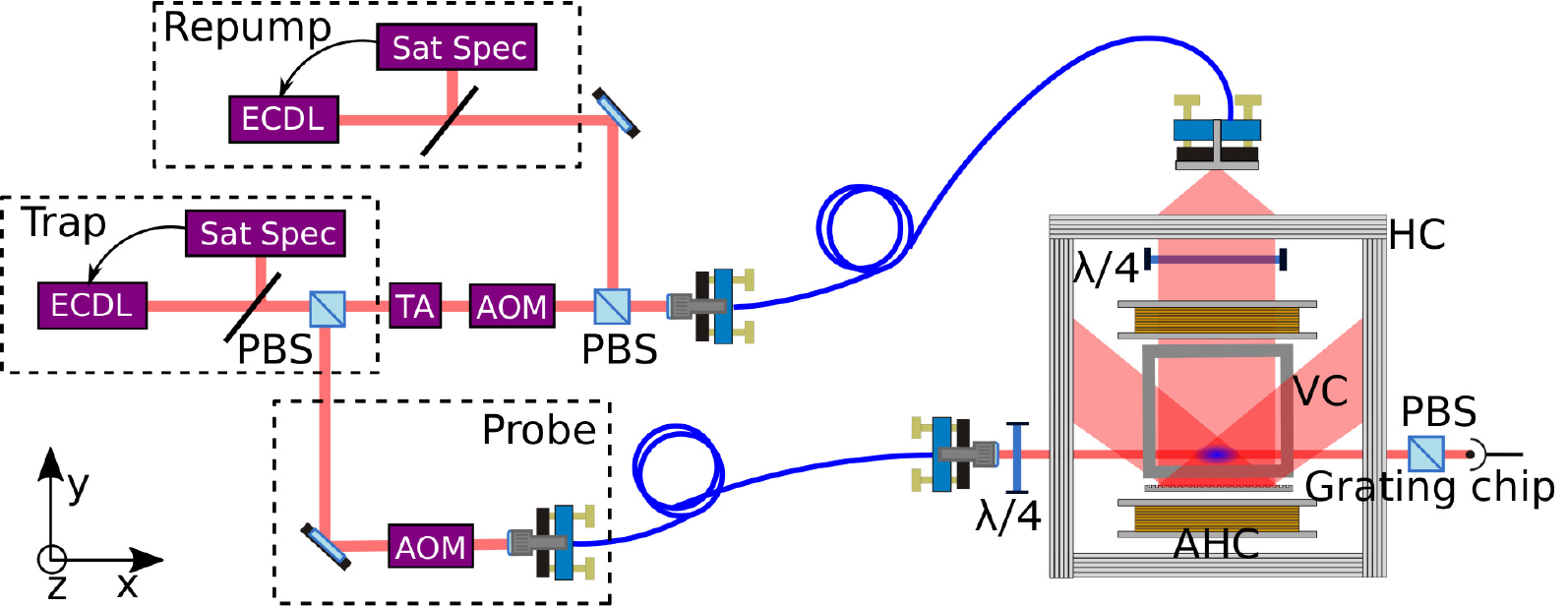}
\caption{Schematic of the grating MOT experimental set-up. ECDL: External cavity diode laser, Sat Spec: Saturated absorption spectroscopy, TA: Tapered amplifier, AOM: Acousto-optical modulator, PBS: Polarising beam splitter, HC: Helmholtz coils, AHC: Anti-Helmholtz coils, VC: Vacuum chamber.}
\label{fig:1}
\end{figure}

The GMOT set-up used here is illustrated in Figure \ref{fig:1}. The trapping and probing light is derived from a single external cavity diode laser, ECDL, to address the D$_2$ transitions of $^{87}$Rb. This laser is locked using saturated absorption spectroscopy and then divided into two separate beams for cooling and pump/probing. The cooling arm is fed into a tapered amplifier (ThorLabs TPA780P20) to provide up to $\approx$300~mW of trap light after fiber coupling. This cooling light is then red detuned by 2$\Gamma$=-12~MHz from the $F$=2 to $F'$=3 transition via an AOM and combined with re-pumping light on a PBS, before being fibre coupled into the chamber. 

The triangular grating is mounted externally from the vacuum system, held horizontally under the glass chamber with the incident light fibre coupled in via a compact tube, which collimates to a 1/e$^2$ diameter of 2~cm and circularly polarises with a quarter wave-plate. Two anti-Helmholtz coils of diameter 12~cm are symmetrically aligned above and below the plane of the grating chip to produce 15~G/cm axial magnetic gradient, with a $25\times25\times12\,$cm Helmholtz coil box used to null stray magnetic fields from the lab environment. Cancellation of these stray magnetic fields has proven critical in achieving a lower temperature optical molasses, resulting in the vacuum system being redesigned with a 2 L/s ion pump to reduce gradient fields in the region of the atomic sample.

The experimental procedure involved a $400\,$ms MOT load directly from background vapour with $10$~mW/cm$^2$, then $10\,$ms of sub-Doppler cooling in a three-stage molasses. This involved detuning the trap light by $30\,$MHz and $60\,$MHz and then halving the intensity in the final stage to achieve temperatures as low as 3.0$\pm$0.7$\,\mu$K, as measured from time of flight imaging \cite{lett1}. This ultra cold ensemble is an ideal starting point for experiments requiring long measurement times and low atomic densities, such as atomic clocks.


\subsection*{Pressure gauge in a compact apparatus}
\label{section:magnetictrap}
To achieve a low temperature molasses in a compact apparatus, we designed our vacuum system with scalability and simplicity in mind, such that a single Rb dispensing getter is used in conjunction with a 2~L/s ion pump (Titan ion pump - Gamma Vacuum). To attain a measurement of the vacuum background pressure at the MOT, we use the relation of the magnetic trap lifetime, $\tau$, to the background Rb pressure in the chamber, $P$, as demonstrated by Monroe \cite{Monroethesis}. The author estimated a lifetime of 10~s for a background pressure at $7\times10^{-10}$~mbar, with the dominant loss contributor being the van-der Waals interaction between the trapped and thermal background atoms. This relation has been utilised in other literature for background pressure measurements at the atoms, each reaching a similar number \cite{pressure1, pressure2}.

The atoms will be magnetically trapped if the magnetic force is larger than that of gravity, where the magnetic force along the gravity direction $z$ can be described mathematically by,
\begin{equation}
F_{z}=-g_{F}\mu_{B}m_{F}\frac{d\bf{B}}{dz},
\end{equation}
Where $g_{F}$ is the Lande g-factor, $\mu_{B}$ is the Bohr magneton and $m_{F}$ is the non-degenerate level of the total angular momentum state. This force can then be used to trap weak-field-seeking atoms if $F_{z}$ is greater than gravity.

\begin{figure}[t]
\centering
\includegraphics[width=10cm]{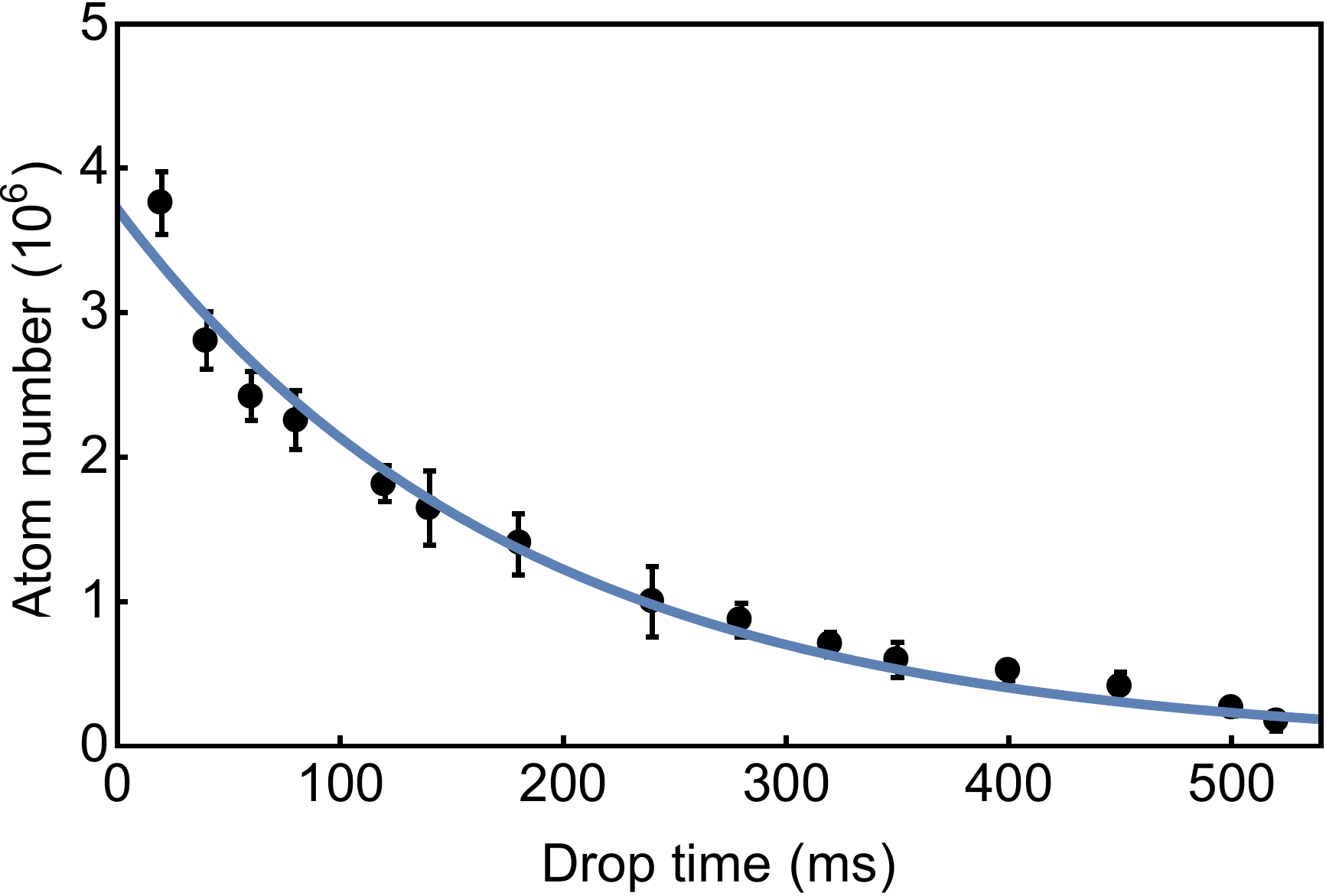}
\caption{Atom number in the magnetic trap as a function of the ballistic drop time. The blue line indicates an exponential fit to determine a 1/e trap lifetime of 180$\pm$10~ms}
\label{fig:magnetictrap}
\end{figure}
The experimental procedure to achieve this begins by trapping the atoms in a quadrupole field, with an axial gradient 15~G/cm. At the end of the molasses sequence the trapping light was turned off and a short optical pumping pulse applied to drive the atoms into the weak-field-seeking $F=2, m_{F}=2$ state. The gradient field from the quadrupole coils was afterwards increased to 50~G/cm, whilst florescence images were taken for a varied time of flight. This enabled the trapped atom number to be described as a function of time, as seen in Figure \ref{fig:magnetictrap}. 

The experimental data points were fit with an exponential decay of the form, $N\exp{(-t/\tau)}$, where $N$ is the initial atom number and $\tau$ is the 1/e magnetic trap lifetime, determined as 180$\pm$10~ms. If we use this $\tau$ to determine the background Rb pressure in the vacuum system, using the relation introduced earlier, we estimate a pressure of 4$\times10^{-8}$~mbar. This pressure is primarily due to the low pumping rate of the scaled down ion pump in an unbaked chamber. However, a $10^{-8}$~mbar pressure is achievable in other systems, where the ion pump has been made redundant in place for specialised micro-fabricated vapour cells with low-He-permeation glass\cite{Argyrios}. Such apparatus would enable future reduction of the vacuum system volume for an unambiguously compact device.

\subsection*{Magnetic sensing for a low temperature molasses}
\label{section:magnetometer}
Optically pumped atomic magnetometers work from the principle of coherent spin precession of a polarised atomic species in the presence of a magnetic field. Larmor spin precession is based on atomic magnetic moments interaction with a local magnetic field, aided by the optical pumping of atoms with near resonant light into a non-degenerate Zeeman level of the ground state via an excited state transition. Once in the stretched state, the angular momentum vector, $\bf{F}$, will precess about the axis of an applied magnetic field, $|\bf{B}|$, at the Larmor frequency, 
\begin{equation}
\omega_{L}=-\gamma |\bf{B}|, 
\label{larmoreqn}
\end{equation}
Where $\omega_{L}$ is the Larmor frequency, $\gamma=\frac{-eg_{F}}{2m}$ is the gyromagnetic ratio of the atom, $e$ is the electron charge and $m$ is the atomic mass.

The precession of $\bf{F}$ around $\bf{B}$ changes the optical absorption and dispersion properties of the atomic ensemble by passing in and out of resonance with the incident polarised electric field. Thus, the precession can be observed as a damped oscillation in the optical absorption and dispersion, measured using a polariser and detector, to give a direct measurement of $|\bf{B}|$ \cite{behbood, Gawlik, Yabuzaki, Yabuzaki2}.

\begin{figure}[t]
\centering
\includegraphics[width=18cm]{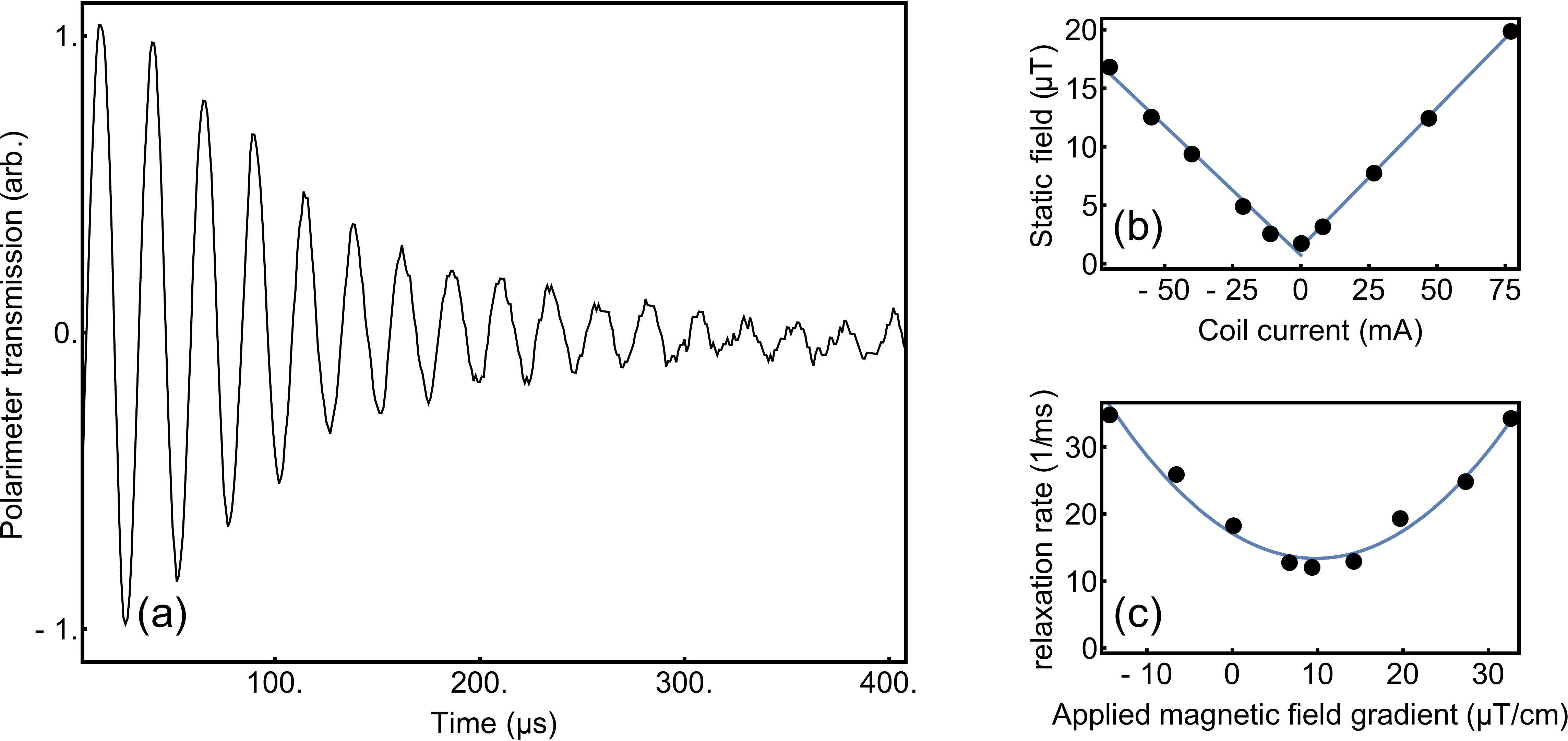}
\caption{(a): Larmor precession measured from the GMOT (b): The static field determined from the Larmor frequency is measured as a function of the Helmholtz coil current with best fitting lines applied. (c): The relaxation rate is measured as a function of applied magnetic gradient field with a best fitting curve.}
\label{fig:larmor}
\end{figure}

Here we measure Larmor precession in an un-shielded grating magneto-optical trap to null stray static fields and reduce the level of gradient fields present during our molasses. The set-up, Figure \ref{fig:1}, uses the same laser source for cooling and probing by splitting the beam into two optical arms. The pump/probe arm is tuned into resonance with the $F$=2 to $F'$=3 level using an AOM. The technique chosen was to use an on-resonant pumping beam, such that a constant intensity beam of 0.68~mW/cm$^2$ and duration of 10~$\mu$s would resonantly pump the atoms with circular light into the stretched state. A 10~$\Gamma$ detuning was then introduced by electronically offsetting the lock-point of the spectroscopy signal for a far off-resonant probe beam. The atomic rotation around an applied magnetic field changes the probing electric field polarisation, allowing a more sensitive detection to be made using a polariser for increased signal-to-noise, as seen in Figure \ref{fig:larmor} (a). This experimental procedure was run in conjunction with a Schmitt trigger from the AC mains in order to reduce AC magnetic field noise by making the magnetic measurement at the same point of the 50~Hz cycle each time.

The precession data is fit with $A\exp{(-t/\tau)}\sin{(2\pi \omega_{L}t+\phi)}$, where $A$ is the initial signal amplitude, $\tau$ is the decay time and $\phi$ is the phase. The Larmor frequency, $\omega_{L}$ is directly related to $|\bf{B}|$ through Eqn. \ref{larmoreqn}, allowing direct correlation of shim current values to the static field in the vicinity of the atomic ensemble. Figure \ref{fig:larmor} (b) illustrates the measurement of the static magnetic field value as a function of the Helmholtz coil current to null the stray fields. This was used on both perpendicular axes to bring the measured field down to the level of the lab noise floor, $\sim$800~nT.

Sensitivity to gradient field components is obtained through the transverse relaxation rate in the ground state coherence, $\gamma=1/\tau$, which is directly related to the total magnetic field gradient through $\gamma=a{|\nabla\bf{B}|}^2+\gamma_{0}$, where $a$ is the coefficient describing the sensitivity to the magnetic field gradient and $\gamma_{0}$ is the relaxation rate in the absence of gradient fields \cite{Budker}. The measured transverse relaxation rates in the ground state coherence are visible in Figure \ref{fig:larmor} (c), where the parameter $\gamma$ is plotted as a function of the difference current between the Helmholtz pair. When applying gradient cancellation on the perpendicular axes by making first order corrections to $d|{\bf{B}}|/dz$ and $d|{\bf{B}}|/dy$ we could improve the transverse decay time from 30~$\mu$s to 100~$\mu$s. A transverse decay time of 100~$\mu$s was the best achievable in our unshielded set-up. 

\begin{figure}[h!]
\centering
\includegraphics[width=10cm]{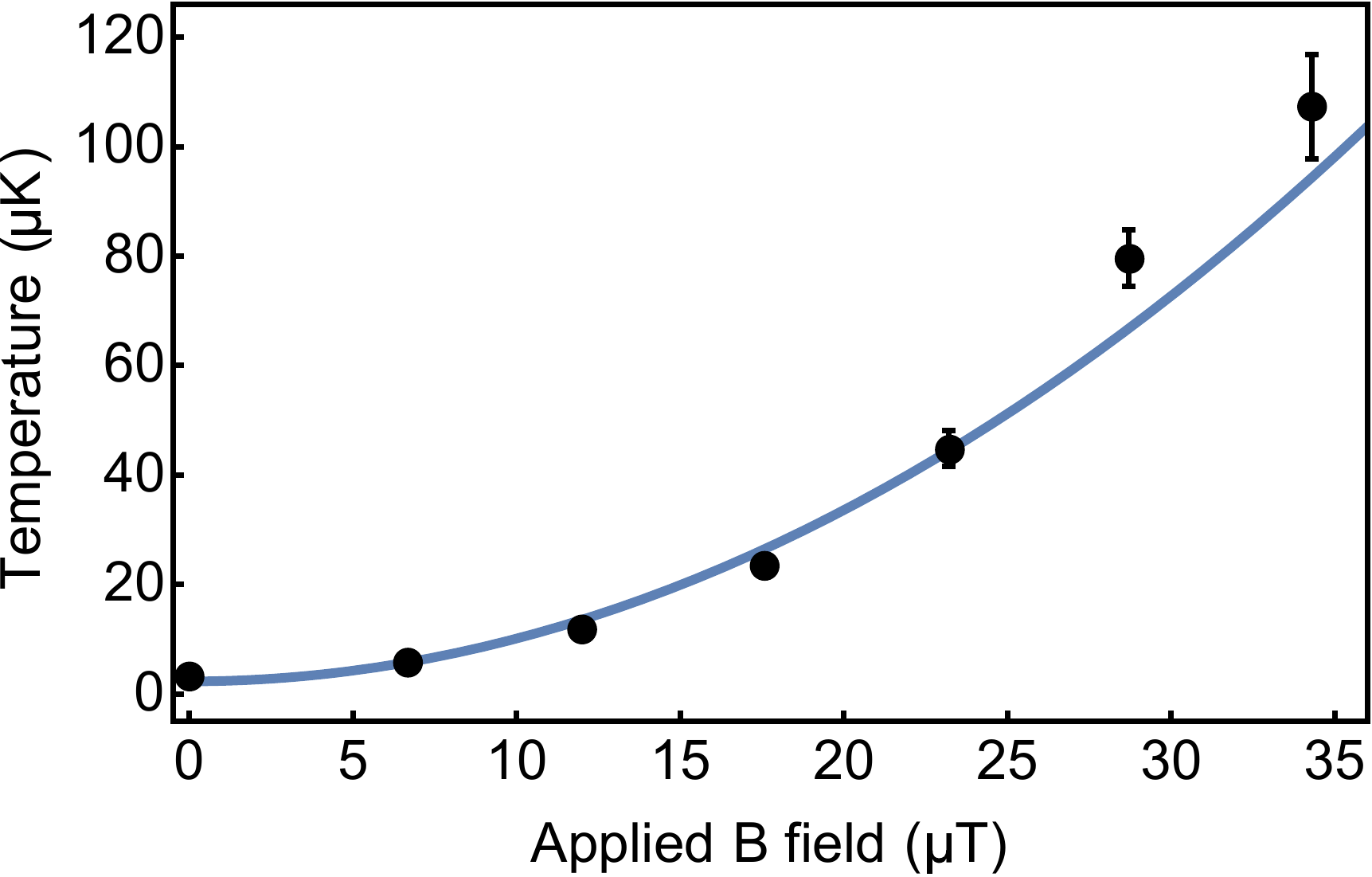}
\caption{Molasses temperature measured as a function of the applied static field with best fitting curve.}
\label{fig:tempvsb}
\end{figure}

With the magnetic gradient fields nulled as well as possible, the temperature in the molasses was then studied as a function of the applied static magnetic field, to demonstrate the necessity for low stray fields when trying to generate ultra-cold atomic samples. As has been discussed by Lett \textit{et al}, the mechanisms of sub-Doppler cooling indicate a sensitivity of temperature to the magnitude of the static magnetic field due to the Zeeman shifts this will induce, destroying the coherent repopulation of $m_{F}$ states involved during polarisation gradient cooling \cite{Lett}. To investigate the level of sensitivity involved in our apparatus we applied static magnetic fields throughout the MOT and molasses stages, measuring the final molasses temperature as a function of the applied static field, as illustrated in Figure \ref{fig:tempvsb}.

The data presented illustrates a temperature range from 3-120~$\mu$K with a 35~$\mu$T change in static magnetic field. The data set is plotted with a quadratic fit, as expected from previous literature \cite{Lett}. We note that our best MOT atom number and molasses temperature combination is ($5\times10^6$, $3.0(7)~\mu$K) - comparable to results in a macro-fabricated grating MOT ($1\times10^6, 7.6(6)~\mu$K) \cite{rolsten1} and an optimised pyramidal MOT system ($4\times10^6, 2.5~\mu$K) \cite{syrte1}.

\subsection*{A cold atom array from a single laser}
Atomic arrays have been utilised in cold atom experiments ranging from quantum information with physical qubit arrays \cite{zollerarray}, to metrological measurements in interferometry and gradiometry \cite{biedermann}. Ionic array apparatus have been miniaturised with trapped ion micro-traps to demonstrate 1D and 2D arrays \cite{monroearray}, however an array of neutral atoms is favourable due to their weaker interaction with the environment \cite{spreeuwarray}. A common method for generating a neutral atom array with a small number/one atom per site is in an optical lattice, where individual microscopic site addressing and measurement is restricted to the regime of the quantum gas microscopes \cite{kuhr1,kuhr2} and optical dipole traps \cite{grangier}.

Large macroscopic arrays of neutral atoms have been achieved in the form of a magnetic lattice loaded from cold atoms, capable of capturing up to 2000 atoms in 500 sites \cite{spreeuwarray}. Macroscopic scaled neutral atom arrays with larger $N$ have also been demonstrated in the form of a 2$\times$2 array of MOTs \cite{pfauarray} where the scalability is limited by the optical set-up. Here we introduce the micro-fabricated grating chip as a simple and robust source for generating 1D and 2D arrays of neutral atoms with a single incident laser and patterned coil geometry. 

\begin{figure}[t]
\centering
\includegraphics[width=12cm]{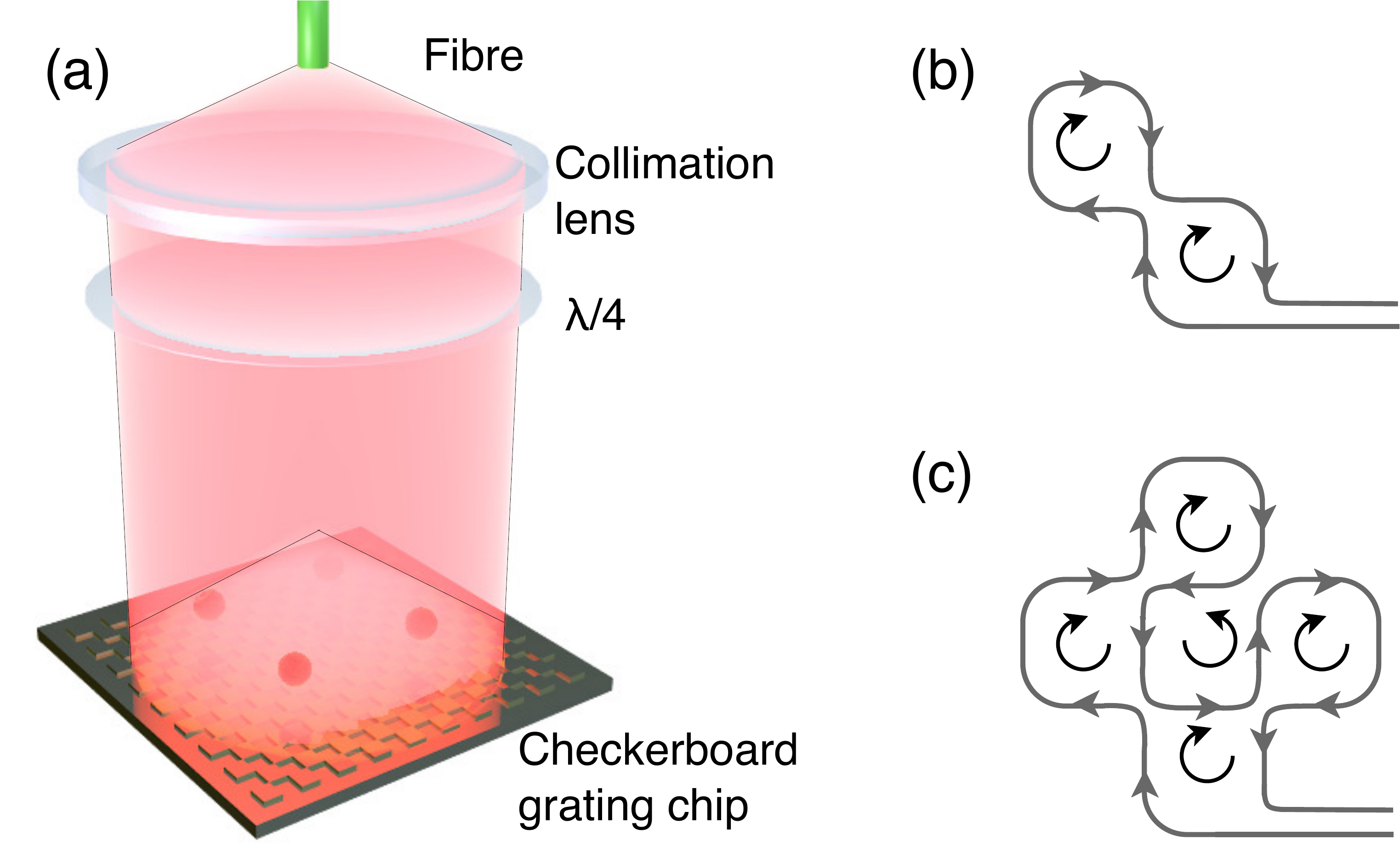}
\caption{(a): Schematic of checkerboard grating. Linearly polarised light diverges from the optical fibre, is collimated and circularly polarised. (b)/(c) Wire pattern for creation of two/four magnetic minima. Grey lines indicate the wire and black curves emphasize the direction of current flow.}
\label{fig:checkerboard}
\end{figure}

For this study, the holographic checkerboard grating chip, discussed in previous literature \cite{nshii,mcgilligan2}, was implemented into the set-up. The benefit of the checkerboard grating is the lack of a central symmetry point that exists in the linear gratings. Instead, the overlap volume extends to fill the surface area of the chip \cite{nshii}, making it accessible to lay a patterned coil beneath the surface of the grating to form a number of MOTs dependent upon the coil geometry, as can be seen in Figure \ref{fig:checkerboard} (a).

To create the appropriate inhomogeneous magnetic potential for multiple MOTs, we used the coil design illustrated in Figure \ref{fig:checkerboard} with an additional external bias field orientated perpendicular to the plane of the coils. The wire geometries used in Figure \ref{fig:checkerboard} (b) and (c) were selected for a 1D and 2D array respectively. Here the grey lines represent the wire formation, with black curves used to emphasize the direction of current flow. For (b), the selected wire formation creates two coils of the same magnitude. In (c), the formation creates five coils, four with the same direction and one in the opposite.

To determine the properties of these wire geometries requires numerical solution of the Biot-Savart law,
\begin{equation}
\textbf{B}(\textbf{r})=\frac{\mu_{0}I}{4\pi}\int \frac{d\textbf{l}\times(\textbf{r}-\textbf{r'})}{|\textbf{r}-\textbf{r'}|^{3}},
\end{equation}
where $\mu_{0}=4\pi\times10^{-7}$~H/m is the permeability of free space, $\textbf{B}(r)$ is the magnetic field at point $\textbf{r}$ from a wire element with current $I$ and unit length $d\textbf{l}$, centred at position $\textbf{r'}$. 

Due to our coils being approximately circular, we assume a perfect circle for our magnetic field calculations. It has been shown \cite{goodelliptic}, that when solved in a cylindrical polar co-ordinate basis, the magnetic field from a circular coil, $\textbf{B}(r,z)$, can be determined for any position $(r,z)$ from the relation of elliptic integrals,
\begin{equation}
\begin{split}
& \textbf{B}_r(r,z)=\frac{\mu_0 I z}{2\pi r}\sqrt{\frac{m}{4\alpha r}}\bigg[\frac{2-m}{2-2m}E-K\bigg],
& \textbf{B}_z(r,z)=\frac{\mu_0 I  }{2\pi r}\sqrt{\frac{m}{4\alpha r}}\bigg[r K+\frac{\alpha m-r(2-m)}{2-2m}E\bigg],
\end{split}
\end{equation}
where $r$ and $z$ are the radial and axial co-ordinates, $\alpha$ is the coil radius, $m=\frac{4\alpha r}{z^2+(\alpha+r)^2}$ and the functions $E$ and $K$ refer to the corresponding elliptic integrals $E(m)=\int_0^{\pi/2}(1-m\sin^2\theta)^{1/2}d\theta$ and $K(m)=\int_0^{\pi/2}\frac{d\theta}{(1-m\sin^2\theta)^{1/2}}$. 

\begin{figure}[t]
\centering
\includegraphics[width=12cm]{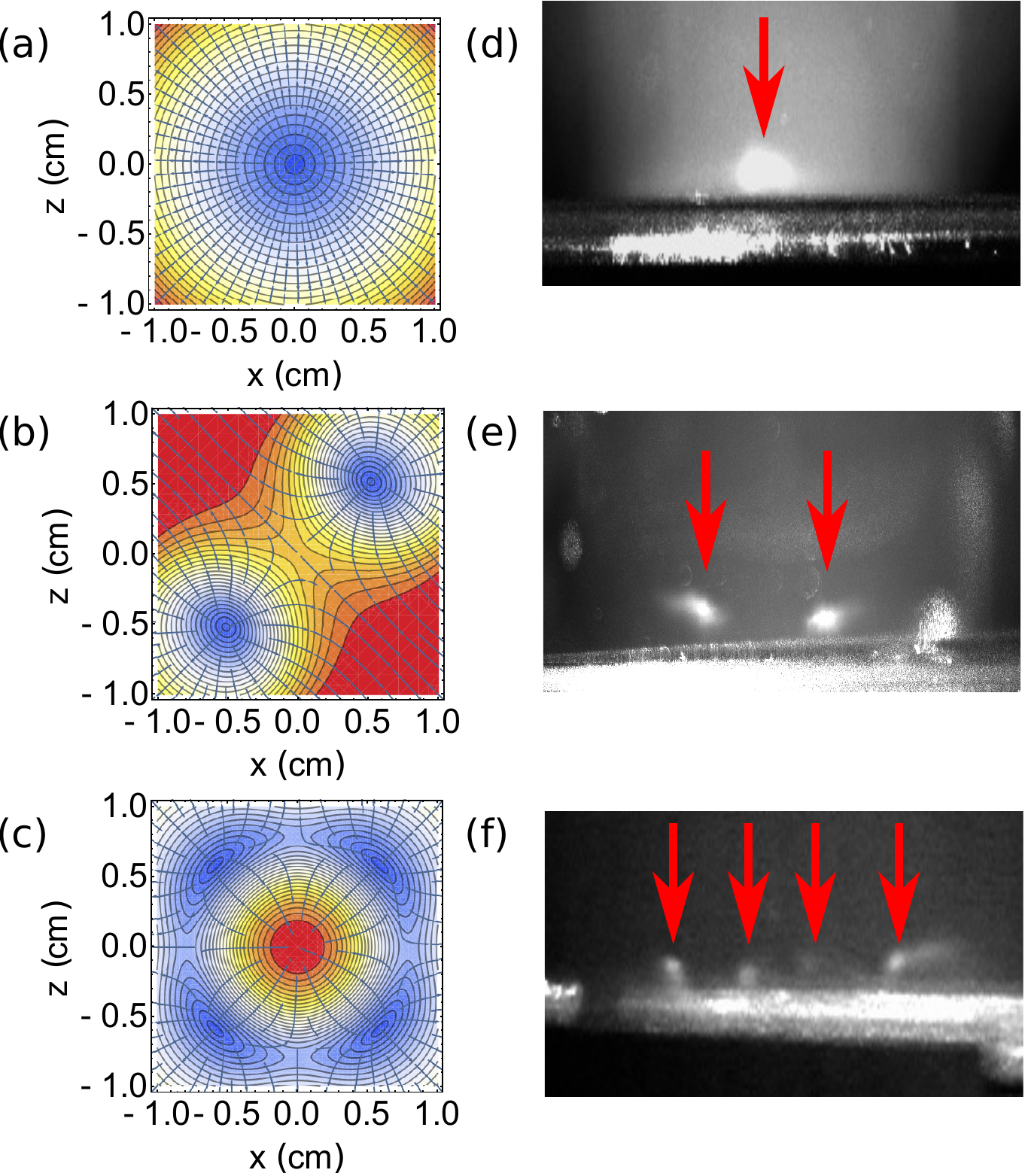}
\caption{(a), (b) and (c): Standard anti-Helmholtz simulation, two magnetic minima simulation for a 1D array and four magnetic minima simulation for a 2D array. (d), (e) and (f): Experimental realisation of a one, two and four MOTs.}
\label{fig:array}
\end{figure}

These equations were used to simulate the magnetic potential for a single MOT, as well for the geometries of Figure \ref{fig:checkerboard} (b) and (c), as shown in Figure \ref{fig:array} (a), (b) and (c) respectively.
The required wire geometries were formed in a 3D-printed mount and placed immediately below the grating. The geometry patterned coil would then be run with a small bias coil in conjunction with an upper coil to create the appropriate magnetic field required. The experimental observations of this can be seen in Figure \ref{fig:array} (d) for a single MOT, (e) for 2 MOTs in a 1D array and (f) for 4 MOTs in a 2D array. The images seen in (d) and (e) were taken with the grating mounted outside of the vacuum, using $\sim$25~mW/cm$^2$ to start with $2\times10^{6}$ atoms in (d) which was then split into 2 MOTs with $5\times 10^{5}$ atoms each. For image (f) the grating and patterned coil were mounted inside the vacuum due to the magnetic minima forming close to the grating surface, forming 4 MOTs using $\sim$2~mW/cm$^2$, with an estimated $<10^5$ atoms in each.

This demonstration proves the potential of the GMOT in future quantum sensing devices requiring an atomic array. Most notably, such a device would prove beneficial to the realisation of a compact gradiometer, where an array of identical MOTs can be used for cancellation of common mode noise.

\section*{Conclusions}
In summary, we have demonstrated ultra-cold temperatures in a grating magneto-optical trap, as a platform for compact quantum sensing. This demonstration looked at the critical parameters for achieving low temperature systems, such as the magnetic environment.

We conclude that pressure measurements determined by the loss rate from a magnetic trap illustrate that these low temperatures can be achieved in a compact vacuum apparatus, even with pressure on the order of 10$^{-7}$~mbar. The introduction of Larmor precessions in a cold $^{87}$Rb atomic sample for the nulling of stray static and gradient magnetic fields provides an accuracy of static field levels below 1~$\mu T$ and an order of magnitude improvement of gradients.

Finally, the demonstration of multiple magneto-optical trap locations from a single incident laser beam coupled to a 2-dimensional diffractive optic opens the door to a compact cold-atom gradiometer. This can be taken further to introduce an array of cold atomic ensembles, with the array number and geometry solely determined by the trapping coil design.

\section*{Acknowledgements}
The authors would like to gratefully acknowledge discussion with C.\ Monroe.
We acknowledge financial support from
the EPSRC (EP/M013294/1), DSTL (DSTLX-100095636R),
and ESA (4000110231/13/NL/PA).

\section*{Author contributions statement}
A.S.A., P.F.G and E.R.\ conceived the experiment(s),  J.P.M. conducted the experiment(s) and wrote the manuscript.  All authors reviewed the results and manuscript.

Additional Information: The authors declare no competing financial interests.



\begin{thebibliography}{99}

\bibitem{katori} Takamoto, M.,  Hong, F.~L., Higashi, R.\ \&  Katori H.\ An optical lattice clock. \doi{10.1038/nature03541}{\textit{Nature} \textbf{435}, 321-324 (2005)}.

\bibitem{ludlow}  Ludlow, A.~D.\ \& Ye, J.\ Progress on the optical lattice clock.\ \doi{10.1016/j.crhy.2015.03.008}{\textit{Comptes Rendus Physique} \textbf{16}, 499-505 (2015)}.

\bibitem{kitching1} Esnault, F.~X., Donley, E.~A., Kitching J.\ \&  Ivanov E.~N. Status of a compact cold-atom CPT frequency standard. \doi{10.1109/FCS.2011.5977843}{Proc.\ 2011 Joint Mtg.\ IEEE Intl.\ Freq.\ Cont.\ Symp.\ and EFTF conf.\ 612-614 (2012)}.

\bibitem{syrte1} Bodart, Q.~\textit{et al}.\ 
A cold atom pyramidal gravimeter with a single laser beam. \doi{10.1063/1.3373917}{\textit{Appl.\ Phys.\ Lett.}, \textbf{96}, 134101 (2010)}.

\bibitem{rolsten1} Lee, J., Grover, J.~A., Orozco, L.~A.\ \& Rolston, S.~L.\ Sub-Doppler cooling of neutral atoms in a grating magneto-optical trap. \doi{10.1364/JOSAB.30.002869}{\textit{J.\ Opt.\ Soc.\ Am.\ B} \textbf{30}, 2869-2874 (2013)}.

\bibitem{zhang} Zhang, H., Li, T. \& Yin, Y.\ Microtrap on a concave grating reflector for atom trapping.\ \doi{10.1088/1674-1056/25/8/087802}{\textit{Chinese Phys. B} \textbf{25}, 087802 (2016)}.

\bibitem{pollock1} Pollock, S., Cotter,  J.~P., Laliotis, A.\ \& Hinds, E.~A.\ Integrated magneto-optical traps on a chip using silicon pyramid structures. \doi{10.1364/OE.17.014109}{\textit{Opt.\ Express} \textbf{17}, 14109-14114 (2009)}.

\bibitem{pyramid1} Camposeo, A.~\textit{et al}.\ 
A cold cesium atomic beam produced out of a pyramidal funnel. \doi{10.1016/S0030-4018(01)01643-1}{\textit{Opt.\ Commun.} \textbf{200}, 231-239 (2001)}.

\bibitem{pyramid2} Arlt, J.~J., Marago, O., Webster, S., Hopkins, S.\ \& Foot, C.~J.\ A pyramidal magneto-optical trap as a source of slow atoms, \doi{10.1016/S0030-4018(98)00499-4}{\textit{Optics Commun.} \textbf{157}, 303-309 (1998)}.

\bibitem{vangeleyn1} 
Vangeleyn, M., Griffin, P.~F., Riis, E.\ \& Arnold, A.~S.\
Single-laser, one beam, tetrahedral magneto-optical trap.\ \doi{10.1364/OE.17.013601}{\textit{Opt.\ Express} \textbf{17}, 13601-13608 (2009)}.

\bibitem{nshii} Nshii, C.~C.~\textit{et al}.\ 
A surface-patterned chip as a strong source of ultracold atoms for quantum technologies. \doi{10.1038/NNANO.2013.47}{\textit{Nature Nanotech.} \textbf{8}, 321-324 (2013)}.

\bibitem{mcgilligan1} McGilligan, J.~P., Griffin, P.~F., Riis, E.\ \& Arnold, A.~S.\ Diffraction grating characterisation for cold-atom experiments. \doi{10.1364/OE.23.008948}{\textit{J.\ Opt.\ Soc.\ Am.\ B} \textbf{33}, 1271-1277 (2016)}.

\bibitem{mcgilligan2} McGilligan, J.~P., Griffin, P.~F., Riis, E.\ \& Arnold, A.~S.\ Phase-space properties of magneto-optical traps utilising micro-fabricated gratings. \doi{10.1364/OE.23.008948}{\textit{Opt.\ Express} \textbf{23}, 8948-8959 (2015)}.

\bibitem{mcgilligan3} Cotter, J. P.~\textit{et al}.\ 
Design and fabrication of diffractive atom chips for laser cooling and trapping. \doi{10.1007/s00340-016-6415-y}{\textit{Appl.\ Phys.\ B} \textbf{122}, 1-6 (2016).}

\bibitem{lett1} Lett, P.~D.~\textit{et al}.\ 
Observation of atoms laser cooled below the Doppler limit. \doi{10.1103/PhysRevLett.61.169}{\textit{Phys.\ Rev.\ Lett.} \textbf{61}, 169-173 (1988)}.


\bibitem{Monroethesis} Monroe, C. \textit{PhD Thesis}, {University of Colorado (1992)}.

\bibitem{pressure1} Weatherill, K.~J.~\textit{et al}.\ 
A versatile and reliably reusable ultrahigh vacuum viewport. \doi{10.1063/1.3075547}{\textit{Rev.\ Sci.\ Instrum.} \textbf{80}, 026105 (2009)}.

\bibitem{pressure2} Monroe, C., Swann, W., Robinson, H.\ \& Wieman, C.\ Very cold trapped atoms in a vapor cell. \doi{10.1103/PhysRevLett.65.1571}{\textit{Phys.\ Rev.\ Lett.} \textbf{65}, 1571-1574 (1990)}.

\bibitem{Argyrios} Dellis, A.~T., Shah, V., Donley, E.~A., Knappe, S.\ \& Kitching, J.\ Low helium permeation cells for atomic microsystems technology. \doi{10.1364/OL.41.002775}{\textit{Opt.\ Lett.} \textbf{41}, 2775-2778 (2016)}.

\bibitem{behbood} Behbood, N.~\textit{et al}.\ 
Real time vector field tracking with a cold-atom magnetometer, \doi{10.1063/1.4803684}{\textit{Appl.\ Phys.\ Lett.} \textbf{102}, 173504 (2013)}.

\bibitem{Gawlik} Sycz, K., Wojciechowski, A.~M.\ \& Gawlik, W.\ Magneto-optical effects and rf magnetic field detection in cold rubidium atoms. \doi{10.1088/1742-6596/497/1/012006}{\textit{J.\ Phys.: Conf. Series} \textbf{497}, 012006 (2014)}.

\bibitem{Yabuzaki} Isayama, T.~\textit{et al}.\ 
Observation of Larmor spin precession of laser-cooled Rb atoms via paramagnetic Faraday rotation.\ \doi{10.1103/PhysRevA.59.4836}{\textit{Phys.\ Rev.\ A} \textbf{59}, 4836-4839 (1999)}.

\bibitem{Yabuzaki2} Takahashi, Y.~\textit{et al}.\ 
Observation of spin echoes of laser-cooled Rb atoms.\ \doi{10.1103/PhysRevA.59.3761}{\textit{Phys.\ Rev.\ A} \textbf{59}, 3761-3765 (1999)}.

\bibitem{Budker} Pustelny, S., Kimball, D.~F., Rochester, S.~M., Yashchuk, V.~V.\ \& Budker, D.\ Influence of magnetic-field inhomogeneity on nonlinear magneto-optical resonances.\ \doi{10.1103/PhysRevA.74.063406}{\textit{Phys.\ Rev.\ A} \textbf{74}, 063406 (2006)}.

\bibitem{Lett} Lett, P.~D.~\textit{et al}.\ 
Optical molasses. \doi{10.1364/JOSAB.6.002084}{\textit{J.\ Opt.\ Soc.\ Am.\ B} \textbf{6}, 2084-2107 (1989)}.

\bibitem{zollerarray} Cirac, J.~I.\ \& Zoller, P.\ A scalable quantum computer with ions in an array of microtraps. \doi{10.1038/35007021}{\textit{Nature} \textbf{404}, 579-581 (2000)}.

\bibitem{biedermann} Rakholia, A.~V., McGuinness, H.~J.\ \& Biedermann, G.~W.\ Dual-axis high-data-rate atom interferometer via cold ensemble exchange. \doi{10.1103/PhysRevApplied.2.054012}{\textit{Phys.\ Rev.\ Applied} \textbf{2}, 054012 (2014)}.

\bibitem{monroearray} Stick, D.~\textit{et al}.\ 
Ion trap in a semiconductor chip. \doi{10.1038/nphys171}{\textit{Nature Phys.} \textbf{2}, 36-39 (2006)}.

\bibitem{spreeuwarray} Whitlock, S., Gerritsma, R., Fernholz, T.\ \& Spreeuw, R.~J.~C.\ Two-dimensional array of microtraps with atomic shift register on a chip. \doi{10.1088/1367-2630/11/2/023021}{\textit{New J.\ Phys.} \textbf{11}, 023021 (2009)}.

\bibitem{kuhr1} Haller, E.~\textit{et al}.\ 
Single-atom imaging of fermions in a quantum-gas microscope. \doi{10.1038/nphys3403}{\textit{Nature Phys.} \textbf{11}, 738 (2015)}.

\bibitem{kuhr2} Sherson, J.~F.~\textit{et al}.\ 
Single-spin addressing in an atomic Mott insulator. \doi{10.1038/nature09827}{\textit{Nature} \textbf{471}, 319-324 (2010)}.

\bibitem{grangier} Bergamini, S.~\textit{et al}.\ 
Holographic generation of microtrap arrays for single atoms by use of a programmable phase modulator. \doi{10.1364/JOSAB.21.001889}{\textit{J. Opt. Soc. Am. B} \textbf{21}, 1889-1894 (2004)}

\bibitem{pfauarray} Grabowski, A.\ \& Pfau, T.\ A lattice of magneto-optical and magnetic traps for cold atoms.\ 
\doi{10.1140/epjd/e2003-00047-3}{\textit{Eur.\ Phys.\ J. D} \textbf{22}, 347-354 (2003)}.

\bibitem{goodelliptic} Good, R.~H. Elliptic integrals, the forgotten functions.\ \doi{10.1088/0143-0807/22/2/303}{\textit{Eur.\ J.\ Phys.} \textbf{22}, 119-126 (2000)}.

\end{thebibliography}
\end{document}